\documentstyle[psfig,aps,prl]{revtex}
\begin{document}
\title{A Quantum-Theoretic Analog for a Pair of Noncommuting
Observables of the Semiclassical Brillouin Function}
\author{Paul B. Slater}
\address{ISBER, University of
California, Santa Barbara, CA 93106-2150\\
e-mail: slater@itp.ucsb.edu,
FAX: (805) 893-7995}

\date{\today}

\draft
\maketitle
\vskip -0.1cm

\begin{abstract}
We study, with the use of {\it numerical}
 integration, a  {\it noncommutative} extension of
 a quantum-theoretic model (an alternative to the semiclassical
 Brillouin function) --- recently
presented by Brody and Hughston and, independently, Slater --- for the
thermodynamic behavior of a spin-${1 \over 2}$ particle. Differences between
the (broadly similar) predictions yielded by this extended model
 and those obtained
 from its conventional (semiclassical/Jaynesian) entropy-maximization
counterpart are examined.
\end{abstract}


\vspace{-0.1cm}
The Brillouin function,
\begin{equation} \label{brill}
- E=\tanh{\beta},
\end{equation}
where $E$ is the expected energy and $\beta$, the inverse temperature
parameter,
 has long served as a model of the thermodynamic behavior of an ensemble
of $N$  noninteracting identical spin-${1 \over 2}$
particles in an applied magnetic field
 \cite{tusy}. In our simplified notation, we take $\beta$ to
represent
 ${\mu B \over k T}$, where $k$ is Boltzmann's constant, $\mu$ is the
particle's magnetic moment, $B$ is the external field strength, $T$
is the temperature, and $\mu B = h$, where $h$ is Planck's constant. So, we
set $h=1$.

Recently, Brody and Hughston \cite{brod} have argued that the theoretical
underpinnings of (\ref{brill}) are semiclassical in nature, since the weighting
of the phase space volume is eliminated, and random phases are averaged.
Park and Band, in an extended series of papers \cite{band},
 questioned the conceptual
foundations of the semiclassical/Jaynesian approach to quantum statistical
thermodynamics (cf. \cite{park1}) --- from which (\ref{brill}) can be derived
\cite[p. 187]{sakurai}. Balian and Balasz \cite{balian} did, however, provide a
rigorous justification (making use of field-theoretic concepts)
 for the Jaynesian (maximum-entropy) method, but, let
us note that their argument
was {\it asymptotic} in nature, relying upon a ``supersystem''
consisting of $N$ replicas of the system, for which they required that
 $N \rightarrow \infty$.

Based on certain metrical considerations, Brody and Hughston
 proposed as a quantum-theoretic
alternative to (\ref{brill}), the function,
\begin{equation} \label{alternative}
- E= {I_{2} (\beta) \over I_{1} (\beta)},
\end{equation}
where the $I$'s represent modified (hyperbolic) Bessel functions.
(This result was also presented --- in a graphical manner --- in a somewhat
earlier paper of Slater \cite{slat1}.) Bessel functions often appear
in the distribution of spherical
 and directional random variables  \cite{robert}.
 Ratios of modified Bessel functions,
such as occur in (\ref{alternative}),
play ``an important role in Bayesian analysis'' \cite{robert}.
It seems important to note, in this regard, the identity,
\begin{equation} \label{stan}
\tanh{\beta} = {I_{{1 \over 2}} (\beta) \over I_{-{1 \over 2}} (\beta)}.
\end{equation}
Lavenda \cite[pp. 193 and 198]{lav} has argued, at considerable length,
that the Brillouin function (\ref{brill}) lacks a suitable {\it probabilistic}
basis, because the integral form for the modified Bessel function
 $I_{\nu}$ exists only for $\nu > {1
\over 2}$.

The relation (\ref{stan}) has been used in expressing
 the expected energy of
 the linear-chain-lattice case ($d=1$)
 of the $D$-vector (or $n$-vector)
model for $D=1$ \cite[p. 492]{stanley}, \cite[p. 370]{robertson},
while the relation (\ref{alternative}) emerges for the instance $D=4$.
The general expression in question takes the form,
\begin{equation} \label{general}
- E = {I_{{D \over 2}}(\beta) \over I_{{D \over 2}-1} (\beta)}.
\end{equation}
The reason that the specific dimension {\it four}, thus, arises in
interpreting the analyses of Brody and Hughston \cite{brod}
and Slater \cite{slat1}, would appear to be
due to the fact that the spin-${1 \over 2}$
states (both mixed and pure) can be considered to lie on the surface
of a hemisphere in {\it four}-dimensional Euclidean space, equipped with
the (natural) Bures or statistical distinguishability
 metric \cite{braun5,uhl,hub1}. Let us also observe that
the (classical) Langevin function \cite{tusy} too is expressible  as a ratio of
modified Bessel functions, that is ($D=3$),
\begin{equation}
\coth{\beta} - {1 \over \beta} =  {I_{{3 \over 2}} (\beta) \over
I_{{1 \over 2}} (\beta)}.
\end{equation}
We also point out that the analysis in \cite{slat1} leads to the result
(cf. (\ref{alternative})) corresponding to $D=6$,
\begin{equation} \label{quatern}
-E = {I_{3} (\beta) \over I_{2} (\beta)},
\end{equation}
for the {\it five}-dimensional convex set (the unit ball in five-space)
 of {\it quaternionic} two-level
systems \cite{adler}.
(The result (\ref{alternative}) is based on the {\it three}-dimensional
convex set --- the ``Bloch sphere'', that is, the unit ball
in three-space --- of the standard {\it complex}
two-level systems, which, as just mentioned above,
 also can be viewed as forming
a hemisphere in four-space \cite{braun5}.)

\begin{figure}
\centerline{\psfig{figure=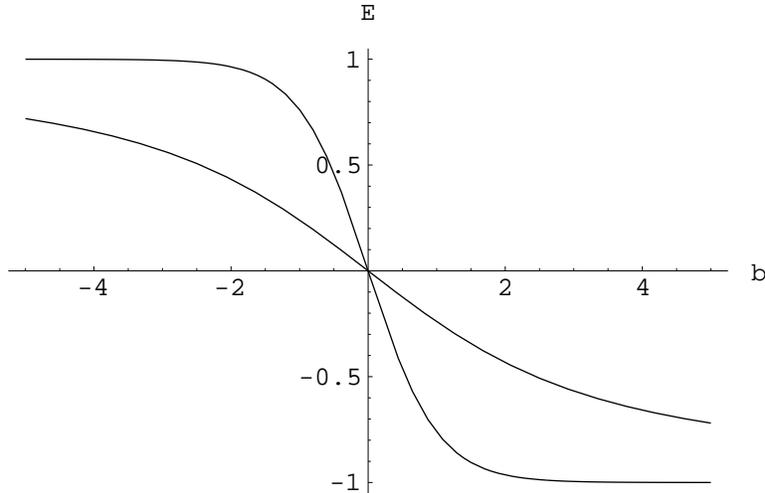}}
\caption{The Brillouin function (\ref{brill}) and the (more steeply-sloped
at $\beta = 0$) quantum-theoretic alternative (\ref{alternative}).
The difference between the two curves is of the greatest magnitude
 (.561292) at $\beta = \pm 1.45489$.}
\label{diff}
\end{figure}

Brody and Hughston have argued that the differences (Fig.~\ref{diff}) in
predictions
 between (\ref{brill}) and
(\ref{alternative}) might be tested in {\it small}
systems (that is, small $N$),
 where ``there seems to be no {\it a priori} reason for adopting
the conventional mixed state approach.''
For an extended discussion of the role of {\it negative} temperatures, in this
context, cf. \cite[sec. 3.52]{robertson}.

Brody and Hughston  noted that the model (\ref{alternative})
 yielded a nonvanishing heat capacity at zero
temperature. ``Since it is known in the case of many bulk substances that
the heat capacity vanishes as zero temperature is approached, it would be
interesting to enquire if a single electron possesses a different
behaviour, as indicated by our results'' \cite{brod}. They also observed
that the increase in magnetization, when the temperature decreases, is slower
for their quantum-theoretic result than for the semiclassical one.

In this letter, we extend the specific
 line of reasoning employed by Slater \cite{slat1} --- based upon the Bures
metric \cite{hub1,hub2,braun1} --- to the case in which, rather than
the expectation value ($E$) of one observable (as in \cite{brod,slat1}),
 one is interested in
fitting the expectation values of {\it two noncommuting}
 observables (cf. \cite{balian,haken}).
We take these observables,
\begin{equation} \label{obs}
\sigma_{1} =  {1 \over 2} \pmatrix{0 & 1 \cr 1 & 0 \cr}, \qquad
\sigma_{2} = {1 \over 2} \pmatrix{0 & -i \cr i & 0 \cr},
\end{equation}
 to be two of the Pauli matrices. (One might also possibly use,
$S_{1} = \sigma_{1}/2, S_{2} = \sigma_{2}/2$, as the spin
observables \cite[p. 38]{bied}.)
 To obtain the conventional
 (semiclassical/Jaynesian) solution to this problem 
\cite{jaynes,larsen}, we express the target density matrix ($\rho$) as
\begin{equation} \label{lars}
\rho = \mbox{exp}{(\Omega \cdot I
 - \lambda_{1} \sigma_{1} -\lambda_{2} \sigma_{2})},
\end{equation}
where $\Omega + 1$ and $\lambda_{i}$ are the Lagrange multipliers for the
normalization of $\rho$ and the measured value of $ \sigma_{i} $,
 respectively.
These multipliers must satisfy
\begin{equation} \label{lag1}
   \Omega = -\mbox{ln} \mbox{Tr} \exp{(-\lambda_{1} \sigma_{1} -\lambda_{2} \sigma_{2})},
\end{equation}
and
\begin{equation} \label{lag2}
{\partial \Omega \over \partial \lambda_{i}} = \langle \sigma_{i} \rangle. 
\qquad (i=1,2)
\end{equation}
The enforcement of these constraints leads to the result,
\begin{equation} \label{brill2}
- \langle \sigma_{i} \rangle =  {\lambda_{i} \tanh{\sqrt{\lambda_{1}^2 +\lambda_{2}^2}} \over
\sqrt{\lambda_{1}^{2} + \lambda_{2}^{2}}}. \qquad (i=1,2)
\end{equation}
Setting either $\lambda_{1}=0$ or $\lambda_{2} =0$, we essentially recover
the Brillouin function (\ref{brill}).

\begin{figure}
\centerline{\psfig{figure=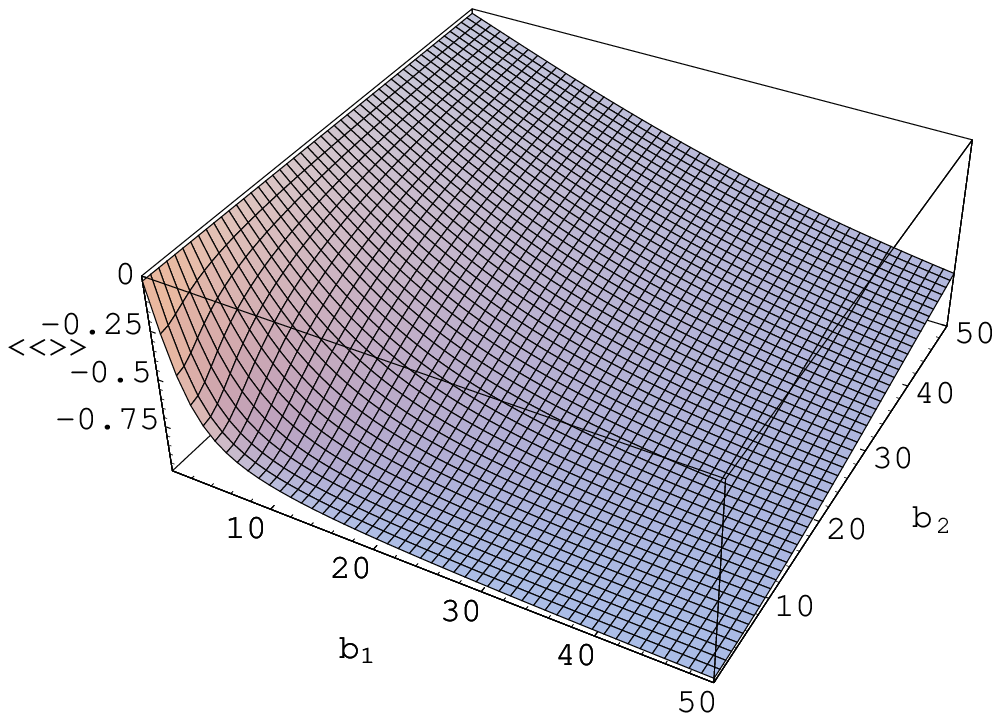}}
\caption{The expected value of the expected value of the observable
 $\sigma_{1}$ as a function of the inverse
temperature parameters, $\beta_{1}$ and $\beta_{2}$,
 of the quantum-theoretic model}
\label{figmean}
\end{figure}

\begin{figure}
\centerline{\psfig{figure=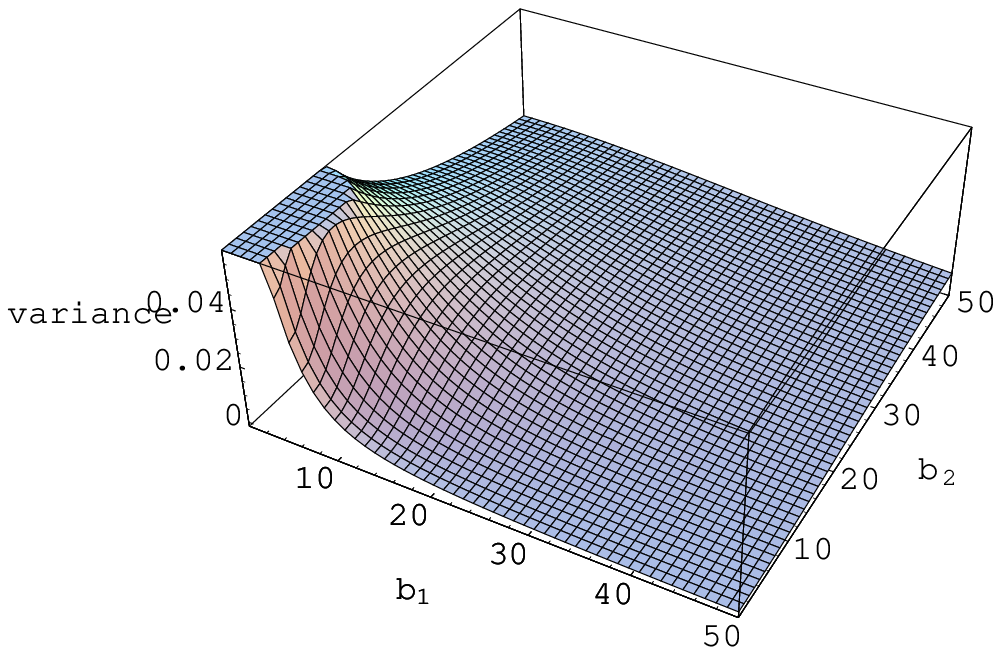}}
\caption{The variance of the expected value of the observable
 $\sigma_{1}$ as a function of the inverse temperature
parameters, $\beta_{1}$ and $\beta_{2}$, of the quantum-theoretic model}
\label{figvar}
\end{figure}

\begin{figure}
\centerline{\psfig{figure=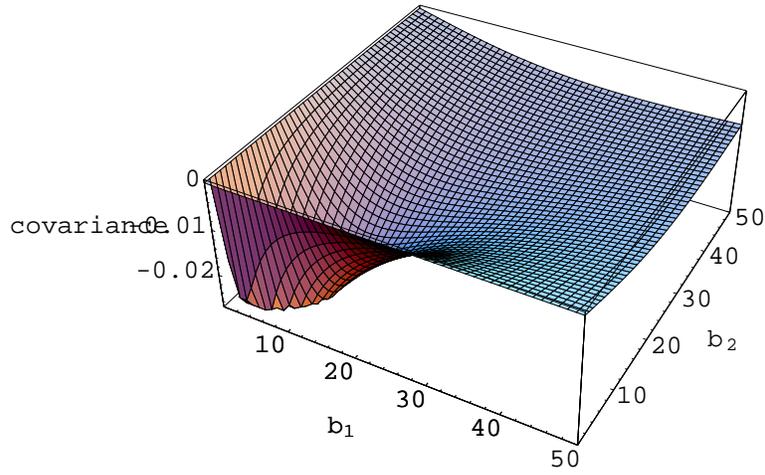}}
\caption{The covariance between the expected values of the observables
 $\sigma_{1}$ and $\sigma_{2}$ as a function of the
inverse temperature parameters, $\beta_{1}$ and $\beta_{2}$,
 of the quantum-theoretic model}
\label{figcovar}
\end{figure}

Now, the volume element of the Bures metric over the three-dimensional convex
set (``Bloch sphere'') of spin-${1 \over 2}$ systems is \cite[eq. (6)]{slat2}
(cf. \cite{slat3,slat4}),
\begin{equation} \label{Bures}
{1 \over 8 \sqrt{1-{\langle \sigma_{1} \rangle}^{2} -{\langle \sigma_{2} \rangle}^{2}
 -{\langle \sigma_{3} \rangle}^{2}}},
\end{equation}
where 
\begin{equation}
\sigma_{3} = {1 \over 2} \pmatrix{1 & 0 \cr 0 & -1 \cr},
\end{equation}
is the additional Pauli matrix.
If we integrate the term (\ref{Bures})
 over one of the three coordinates (say, $\langle \sigma_{3}
 \rangle$),
we obtain simply a {\it uniform} distribution ($\pi /8$) over the unit disk
($0 \leq {\langle \sigma_{1} \rangle}^{2} +{\langle \sigma_{2} \rangle}^{2} \leq 1$).
Interpreting this uniform distribution as a density-of-states or structure
function, we can apply a bivariate Boltzmann factor, $e^{-\beta_{1}
\langle \sigma_{1} \rangle -\beta_{2} \langle \sigma_{2} \rangle}$,
 to it. Integrating this product over the
$\langle \sigma_{2} \rangle$-coordinate (between the limits of $\pm
\sqrt{1- {\langle \sigma_{1} \rangle}^{2}}$), we obtain,
\begin{equation} \label{marginal}
{e^{- \beta_{1} \langle \sigma_{1} \rangle} \pi \sinh{(\beta_{2}
\sqrt{1 - {\langle \sigma_{1} \rangle}^2})} \over 4 \beta_{2}}.
\end{equation}
The corresponding partition function  is the integral of (\ref{marginal})
over the remaining coordinate ($\langle \sigma_{1} \rangle$) from -1 to 1.
This integration has to be performed {\it numerically}.
Carrying this out, we are  then able, again employing numerical
integration,
to obtain the (twofold)
 expected value ($\langle \langle \sigma_{1} \rangle \rangle$)
 of $\langle \sigma_{1} \rangle$ (Fig.~\ref{figmean})
as a function of $\beta_{1}$ and $\beta_{2}$, as well as the variance
about this expected value (Fig.~\ref{figvar}), and the covariance between
$\langle \sigma_{1} \rangle$ and $\langle \sigma_{2} \rangle$ (Fig.~\ref{figcovar}).
(The covariance is the expected value --- with respect to the Boltzmann
distribution --- of the product $(\langle \sigma_{1}
\rangle - \langle \langle \sigma_{1} \rangle \rangle)
(\langle \sigma_{2} \rangle - \langle \langle \sigma_{2} \rangle \rangle)$.)

\newpage

\vspace{-1.25in}
\vspace{-.4in}
\begin{figure}
\centerline{\psfig{figure=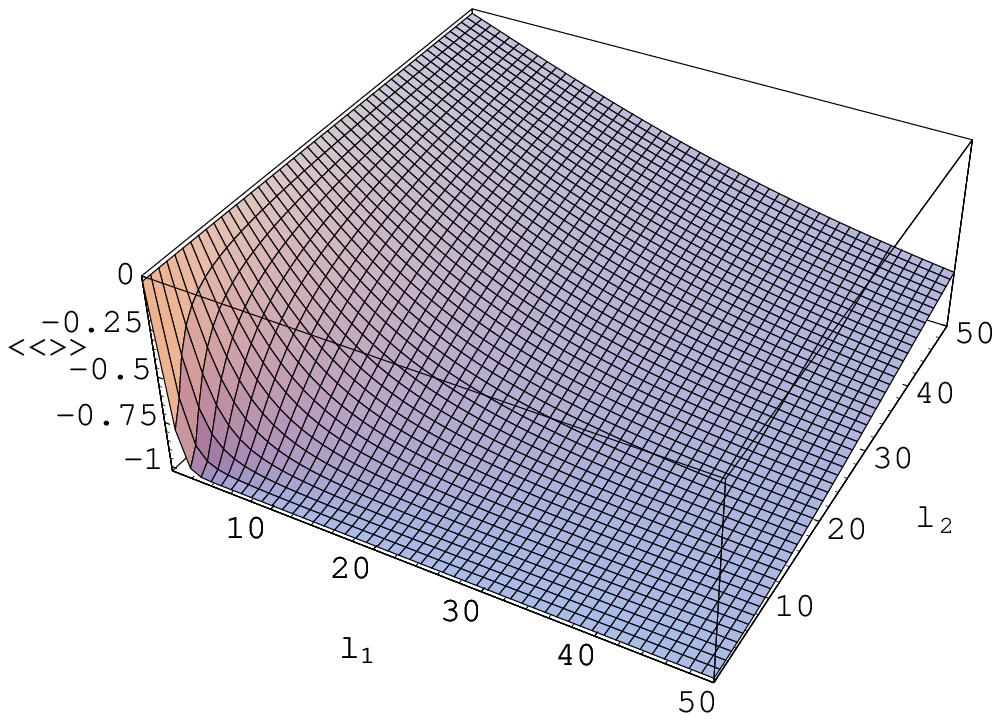}}
\caption{The expected value of the expected value of the  observable
$\sigma_{1}$ as a function of the inverse temperature parameters, $\lambda_{1}$
and $\lambda_{2}$, of the
semiclassical (Brillouin-type) model}
\label{larsenmean}
\end{figure}
\vspace{-.4in}
\begin{figure}
\centerline{\psfig{figure=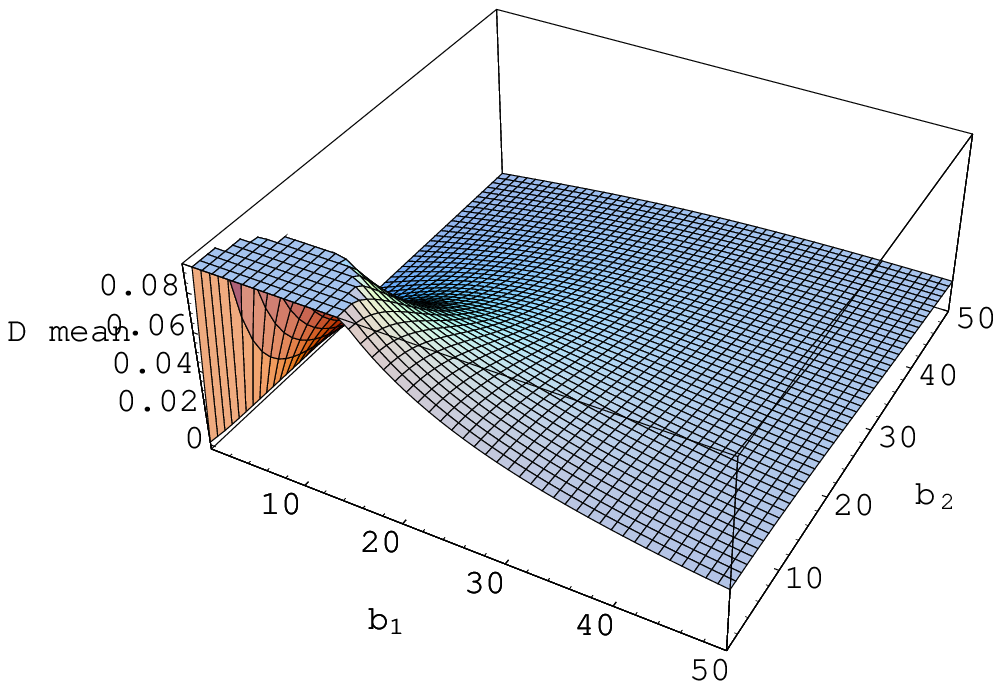}}
\caption{The difference between the quantum-theoretic results 
(Fig.~\ref{figmean}) and the
semiclassical ones (Fig.~\ref{larsenmean})
 for the expected value of the expected value of the
 observable $\sigma_{1}$ --- having identified the $\lambda$'s with the
corresponding $\beta$'s}
\label{difm}
\end{figure}
\begin{figure}
\centerline{\psfig{figure=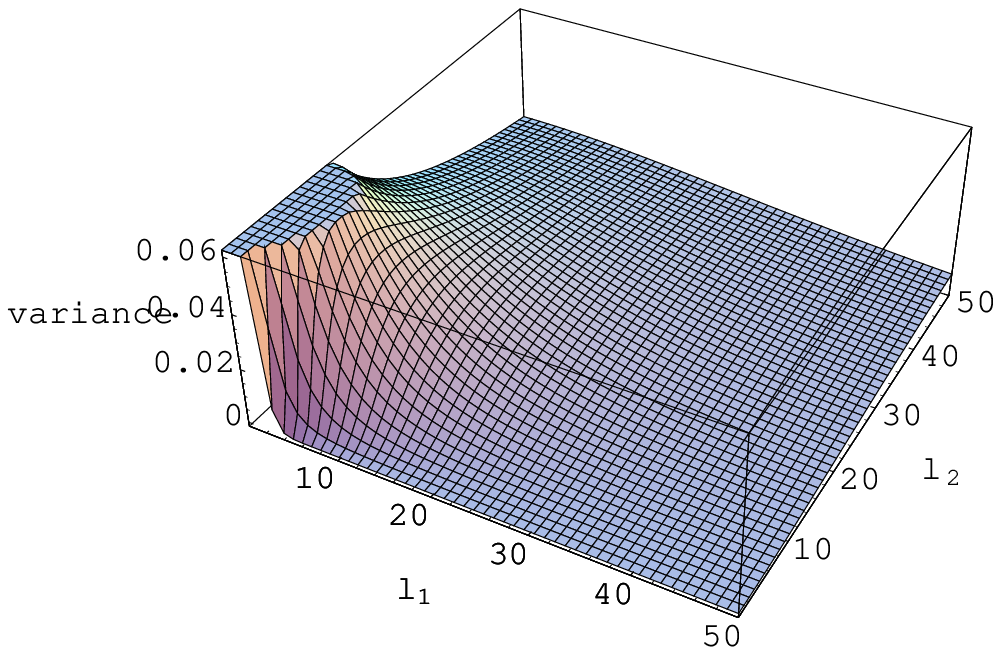}}
\caption{The variance of the expected value of the observable $\sigma_{1}$ as a
function of the inverse temperature parameters, $\lambda_{1}$ and
$\lambda_{2}$, of the semiclassical (Brillouin-type)
model}
\label{larsenvar}
\end{figure}
\begin{figure}
\centerline{\psfig{figure=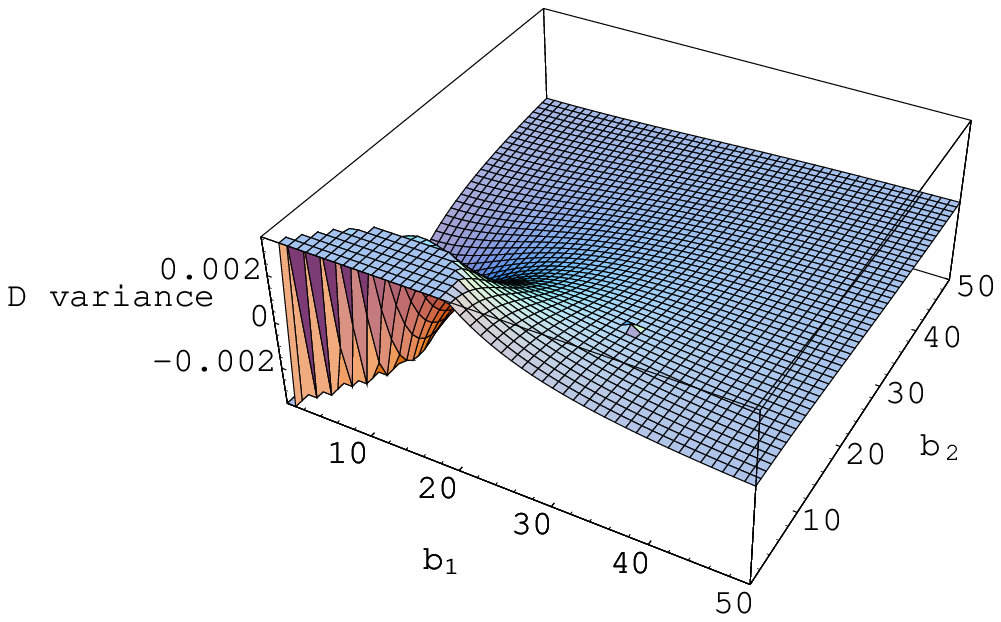}}
\caption{The difference between the quantum-theoretic results
(Fig.~\ref{figvar}) and the
semiclassical ones (Fig.~\ref{larsenvar})
 for the variance of the expected value of the
observable $\sigma_{1}$ --- having identified the $\lambda$'s with the
corresponding $\beta$'s}
\label{difv}
\end{figure}
\begin{figure}
\centerline{\psfig{figure=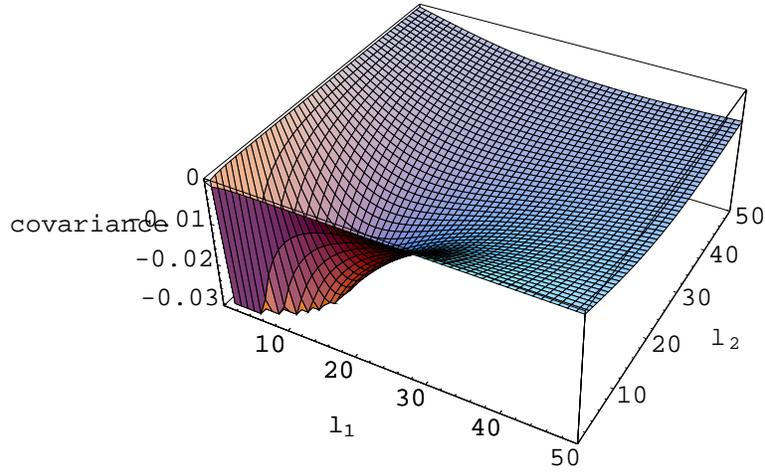}}
\caption{The covariance between the expected values of the observables
$\sigma_{1}$ and $\sigma_{2}$ as a function of the inverse temperature
parameters, $\lambda_{1}$ and $\lambda_{2}$, of the semiclassical
(Brillouin-type) model}
\label{larsencovar}
\end{figure}
\begin{figure}
\centerline{\psfig{figure=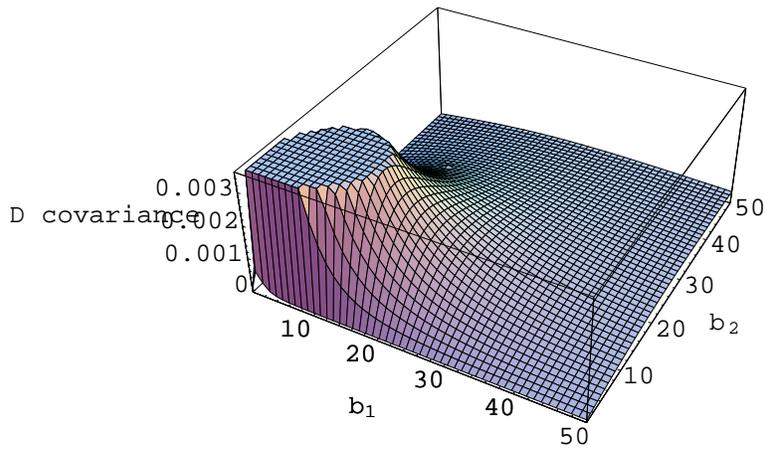}}
\caption{The difference between the quantum-theoretic results
 (Fig.~\ref{figcovar}) and the
semiclassical ones (Fig.~\ref{larsencovar})
 for the covariance of the expected values of the
observables $\sigma_{1}$ and $\sigma_{2}$--- having identified the $\lambda$'s with
the corresponding $\beta$'s}
\label{difc}
\end{figure}

For comparison purposes (cf. Fig.~\ref{diff}),
 we present the semiclassical (noncommuting Brillouin)
counterparts to these quantum-theoretic results in the companion figures
 (Figs.~\ref{larsenmean}, 
\ref{larsenvar}, \ref{larsencovar}). We note strong qualitative resemblances
between the two members of each of these
 three pairs of figures.  We also present in Figs.~\ref{difm},
\ref{difv} and \ref{difc}, the differences obtained (after setting
$\lambda_{i}$ to $\beta_{i}$, $(i=1,2)$) by subtracting the semiclassical
results (shown in Figs.~\ref{larsenmean}, \ref{larsenvar} and
 \ref{larsencovar})
from the corresponding quantum-theoretic ones (given in Figs.~\ref{figmean},
\ref{figvar} and \ref{figcovar}). The most substantial
 differences in all three cases
appear in the vicinity of the (high-temperature)
 origin ($\beta_{1} = \beta_{2} =0$).

Let us, in conclusion, consider the possibility of expanding the analyses
above to the case of {\it three} noncommuting observables, rather than two.
Then, we would apply a trivariate Boltzmann factor,
$e^{-\beta_{1} \langle \sigma_{1} \rangle -\beta_{2} \langle \sigma_{2} \rangle -
\beta_{3} \langle \sigma_{3} \rangle}$, to the volume element
(\ref{Bures}) itself of the Bures metric (rather than its two-dimensional
[uniform]
marginal --- ${\pi \over 8}$). Integrating out the $\langle \sigma_{3} \rangle$-coordinate
(between the limits $\pm \sqrt{1 - {\langle \sigma_{1} \rangle}^2
-{\langle \sigma_{2} \rangle}^2}$), we obtain (cf. (\ref{marginal})),
\begin{equation} \label{3Boltz}
{\pi e^{- \beta_{1} \langle \sigma_{1} \rangle - \beta_{2} \langle \sigma_{2} \rangle}
J_{0}(\beta_{3} \sqrt{{\langle \sigma_{1} \rangle}^2 + {\langle \sigma_{2} \rangle}^2})
\over 8},
\end{equation}
where $J_{0}$ is a Bessel function of the first kind. To obtain the
corresponding partition function, it then appears necessary, similarly to
before,  to 
{\it numerically} integrate (\ref{3Boltz}) over the unit disk
$(0 \leq {\langle \sigma_{1} \rangle}^2 + {\langle \sigma_{2} \rangle}^2 \leq 1)$.

\acknowledgments

I would like to express appreciation to the Institute for Theoretical
Physics for computational support in this research.

\end{document}